\documentclass[12pt]{article}
\begin{document}
\vspace*{4cm}

\begin{center}
{\Large \bf How Can We Go From Hadron Collider Data Toward the Underlying
Theory That Extends the Standard Model? After the
Champagne...}\\
\vspace{.2in}
Gordon Kane\\
Michigan Center for Theoretical Physics\\
University of Michigan\\
Ann Arbor MI 48109
\end{center}

\begin{abstract}
This is a composite based on talks at Physics at LHC,
Vienna, July 2004, TeV4LHC, Fermilab, Sept. 2004, and the String
Phenomenology meeting, Perimeter Institute, March 2005.
\end{abstract}

\section{Introduction} The main goal of particle physics now must be to
learn experimentally what theory extends and strengthens the foundations
of the Standard Model (SM) of particle physics. While major clues can
come from indirect information such as rare decays, magnetic dipole
moments, proton decay, and particularly wimp detection in dark matter
detection experiments (which could essentially establish the existence
of superpartners) and electric dipole moments that probe the CP
structure, or even from theory, it will be crucial to directly detect
(or not) superpartners at colliders before we are sure of nature's
answer. This could happen at the Tevatron collider soon, and must happen
at the LHC if the Supersymmetric Standard Model (SSM) is indeed the
correct approach (as is implied by several indirect arguments and
particularly by LEP data).

Once the signal beyond the Standard Model is found, we will celebrate.
Then what? How do we test that what is discovered is indeed
supersymmetry?  We probably know how to do that, from various signals
and their patterns.  To make progress toward understanding the world,
and to move toward knowing the theory that underlies the
Supersymmetric Standard Model is it enough to only know that nature is
supersymmetric, or do we have to learn about how supersymmetry is
broken and about the underlying (presumably) string theory? This is
analogous to asking if we needed LEP to study the Standard Model and
learn about how to extend it. But we are not well prepared to learn
from hadron collider data.

To demonstrate we are seeing superpartners, and to learn if string
theory is relevant to our world presumably we have to learn a great
deal about the superpartners and their properties and interactions. In
this talk I will assume the discovery will in fact be of supersymmetry
-- the kinds of issues I raise will apply whatever is found. I will
discuss how we can make progress in learning about how supersymmetry
is broken, and about the underlying theory. For simplicity I will also
assume the underlying theory is a string theory and not keep making
qualifying remarks.

It could happen, if we are lucky, that a very simple pattern will
emerge, with only a few light superpartners, or a highly degenerate
one such as mSUGRA (with degenerate gaugino masses, degenerate
diagonal sfermions and trilinear's, and all masses real). Such
approaches are easy to test. Alternatively, though string theories
cannot yet provide convincing low scale predictions, they may be
suggestive of qualitatively what to expect, in which case few
degeneracies and complex masses are rather likely. It is this kind of
world we want to prepare to study.

Much useful work has been done to learn how to measure properties of new
physics at LHC. In this talk I will emphasize obstacles to determining
the relevant properties in the general case, concerning which much less
work has been done. Here I will only provide a few references from which
the literature can be traced. More complete citations will be provided
elsewhere. A recent review \cite{A} covers a number of relevant topics,
and two papers \cite{B,E} focus on the approach described in this talk,
both with many references.

\section{Obstacles}In fact, there are a number of obstacles to learning
about such issues as supersymmetry breaking and the underlying
theory. Two problems are low-scale ones.

\bigskip

I. The first major problem that arises is that experiments
measure cross sections times branching ratios, $\sigma \times BR,$ and
masses of mass eigenstates (actually, mostly not even the mass
eigenstate masses but mass-squared differences). None of these are in
theorist's Lagrangians.  The gluino mass doesn't mix with others in a
mass matrix, but it gets major SQCD radiative corrections. Chargino
and neutralino and stop masses are only related to the (complex)
soft-breaking parameters $M_{1},$ $M_{2},$ $ \mu ,$ and the soft stop
left- and right-handed masses, for example, via being eigenvalues of
mass matrices. Experimental talks on discoveries at the Tevatron or
LHC will report $\sigma \times BR$ in $fb$ units, and kinematical
distributions --- learning the implications of such quantities is not
straightforward.

\bigskip

II. If one then studies how one can learn the Lagrangian parameters
from what is measured, one finds the second major problem that at
hadron colliders there are always more Lagrangian parameters than
observables, so one cannot in general solve for the Lagrangian
parameters. This is actually the most important reason why we need a
linear electron collider, but that is not our subject --- since it is
very unlikely a linear collider will exist until well over a decade
after LHC reports data, we will focus entirely on hardon collider
physics here. So there is no general way known to measure $ \tan \beta
,$ $\mu ,$ gaugino masses, scalar masses, trilinear couplings,
etc. But these are the quantities we need to know to learn about
supersymmetry breaking and the underlying theory. So we have to invent
techniques to try to determine these quantities as best we can. The
main approach that has been used historically is to make ad hoc
assumptions until there are fewer parameters than observables and then
``solve'' for the parameters, but unless we are lucky and the people
making assumptions are guessing right, this approach may not give the
right answers.

We may be lucky also if nature is in some easy corner of parameter space,
e.g. if some process such as $B_{s}\rightarrow \mu ^{+}\mu ^{-}$ is observed
that requires large $\tan \beta ,$ or if prompt photons are observed in
supersymmetry events. Then some properties of the theory and the
parameters can be deduced, and the data will be less obscure. But in
general we need to develop new methods. It may be important to do that
before there is data, since the methods may work better for some trigger and
analysis techniques than others.

This situation with mass eigenstates and Lagrangian parameters may not be
familiar to all readers. We can see how it works in the simplest example,
the charginos. The chargino mass matrix follows from the $L_{soft},$ which
contains the complex gaugino mass parameters $M_{2},$ and $\mu ,$ and
depends on $\tan \beta ,$ in the wino-higgsino basis (where we
explicitly show the phases):

\begin{eqnarray}
M_{\widetilde{C}}=\left( 
\begin{array}{cc}
M_{2}e^{i\phi _{2}} & \sqrt{2}M_{W}\sin \beta  \\ 
\sqrt{2}M_{W}\cos \beta  & \mu e^{i\phi _{\mu }}
\end{array}
\right).  
\end{eqnarray}

The off-diagonal elements can be complex too since they arise when the
Higgs gets a vev, and the Higgs vev can be complex, but I will just keep
the phases of $M_{2}$ and $\mu $ here to illustrate in the simplest
reasonable case.The masses of the mass eigenstates are the eigenvalues
of this matrix. To diagonalize it one can form the hermitean matrix
$M^{\dag }M.$ The easiest way to see the main points are to write the
sums and products of the mass eigenstates,

\begin{equation}
\begin{array}{rcl}
M_{\widetilde{C_{1}}}^{2}+M_{\widetilde{C}_{2}}^{2}=TrM_{\widetilde{C}
}^{\dag }M_{\widetilde{C}}=M_{2}^{2}+\mu ^{2}+2M_{W}^{2},\hspace{1in} & & \\ 
M_{\widetilde{C}_{1}}^{2}M_{\widetilde{C}_{2}}^{2} =\det
M_{\widetilde{C} }^{\dag }M_{\widetilde{C}}=M_{2}^{2}\mu
^{2}+2M_{W}^{4}\sin ^{2}2\beta -2M_{W}^{2}M_{2}\mu \sin 2\beta \cos
(\phi _{2}+\phi _{\mu })\\ 
\end{array}\label{eq:spa}
\end{equation}

Experimenters measure the masses of the mass eigenstates (sometimes
only mass differences). One thing to note is that the masses depend on
the phases $ \phi _{2}$ and $\phi _{\mu },$ even though there is no CP
violation associated with the masses. Often it is implicitly assumed
that phases can only modify CP-violating effects, but that is not so.
The combination $\phi _{2}+\phi _{\mu }$ is a physical phase,
invariant under any reparameterization of phases, as much a basic
parameter as $\tan \beta $ or any soft mass. (This parameter is
constrained by electric dipole moment data, though not determined, but
a number of other phases are less constrained.) There are 4
parameters, the magnitudes of $M_{2}$ and $\mu ,$ the physical phase,
and $\tan \beta .$ Measuring the two mass eigenstates is not enough to
determine 4 parameters. Adding cross section and decay data will help,
but then squark and slepton masses and couplings enter, and the number
of parameters does not converge as one adds sectors. There are
normally too few observables. Perhaps the phase is small, but so far
no argument has been given that might explain why it is small, so it
should be measured.

One can go to the Higgs sector. There it is necessary to add the one loop
contributions (at tree level $m_{h}<M_{Z}),$ and the minimum number of
significant parameters is seven \cite{C}, so at least seven observables are
needed. It may be possible to improve the situation by combining the Higgs
and gaugino sectors, since some parameters are common, but no general study
has been done yet.

\bigskip

III. There are additional obstacles, of which we mention three
categories (two have to do with connecting low and high scales, the
third is relevant at all scales):

(a) If one tries to do top-down calculations, one has to know the
whole theory before observables can be predicted. The superpotential
$\mu $ parameter must be zero before supersymmetry breaking, but it is
not generated in the same way as the soft parameters.  There are
approaches to calculating $\mu ,$ but in practice there are as yet no
examples of generating $\mu$ and $\tan \beta$ theoretically in models
and satisfying the EWSB conditions without forcing them. For $\tan
\beta $, which is extremely important in practice since it appears in
essentially all testable predictions, the problem is worse since it
does not exist in the high scale theory. It is only generated after
electroweak symmetry breaking. Before we can say we have an
understanding of any underlying approach we must be able to calculate
$\tan \beta $ and deduce EWSB but this has never been done. So it is
very unlikely any predictive top-down calculations can be done. As
remarked above, there is no general way known to measure $\tan \beta$.

(b) The renormalization group equations used to connect high and low
scale data are extremely model dependent. They change significantly if
intermediate scale matter is included. The results depend on what low
and high scales are chosen to begin and end the running. For some
quantities infrared fixed points obscure the running. Much more study
is needed to understand how to test for ambiguities and uniqueness.

(c) The Lagrangian parameters are complex. The phases significantly
affect not only CP violation, but superpartner masses, production and
decay rates, the higgs sector, dark matter relic density and detection,
and more.  No known symmetry or argument implies the phases should be
small. There are some constraints on some of them, but many could be
large.

All of these obstacles could be solved with sufficient precision data,
but that will not be available for over a decade at best. They can
probably be solved anyhow with good analysis and techniques that could
be developed in the next few years if people think hard.

\section{Two Approaches --- Consistency, and Signature Space Projections}

There are two ways that may help to bypass the obstacles. First,
supersymmetry is a real theory. Everything is known but the (complex,
non-flavor-diagonal) masses and the vevs (just as in the SM). If one
tries to find a set of masses that reproduce a particular signal, there
are actually many subtle constraints since different processes involving
the same masses must be consistent. Production cross sections limit
other masses. Some decays must occur and others must not, which further
constrains masses. So even with one solid signal the possible Lagrangian
parameters are remarkably constrained. Then further data from other
processes can add major information. If we understand the theory well
enough, and do not make misleading assumptions, we may be able to make
considerable progress.

Second, and potentially very important, patterns of \textit{Inclusive
Signatures} may be very powerful \cite{D}. An inclusive signature is
anything that can really be measured. The table illustrates this.
Columns describe different ways of breaking supersymmetry, different
underlying theories. Each row is a different inclusive signature. All
collider ones have missing transverse energy. No single signature will
tell us much about supersymmetry breaking or the underlying theory,
but the pattern of several will. The number of signatures can be
extended, and should include collider and non-collider ones. The
number of columns can be greatly extended as more theories are
studied. Our beginning studies \cite{B,E} suggest that such patterns
are very powerful at distinguishing underlying physics even though
assumptions are necessarily involved to relate theory and data. As
studies improve and there is data we expect this signature space
approach to be of great value, perhaps allowing the possibilities of
testing specific susy-breaking approaches, testing gaugino mass
unification even though separate gaugino masses cannot be measured,
determining $\tan \beta$, etc.

\begin{center}
\begin{table}[ph]
{\footnotesize
\begin{tabular}{@{}c|rrrr@{}}
{}&{}&{}&{}&{}\\[1ex]
Inclusive Signatures & $\widetilde{G}MSB,$ large $\mu $ &
$\widetilde{G}MSB,$
small $\mu $ & $GMSB$ & dilaton dominated\\[1ex]
\hline
same sign dileptons & yes\hspace{1cm} & yes\hspace{1cm} &
yes\hspace{.5cm} & yes\hspace{1.5cm} \\[1ex] 
prompt $\gamma ^{\prime }s$ & no\hspace{1cm} & sometimes\hspace{.5cm} &
yes, but \hspace{.5cm}& no\hspace{1.5cm} \\[1ex] 
trileptons & yes\hspace{1cm} & no\hspace{1cm} & no\hspace{.5cm} &
yes\hspace{1.5cm} \\[1ex] 
very large missing E$_{T}$ & {} & {} & {} & {}\\[1ex] 
b-rich & {} & {} & {} & {}\\[1ex] 
opposite sign dileptons & {} & {} & {} & {}\\[1ex] 
$B_{s}\rightarrow \mu ^{+}\mu ^{-}$ & {} & {} & {} & {}\\[1ex] 
direct wimp detection & {} & {} & {} & {}\\[1ex] 
P$_{T}$ end points & {} & {} & {} & {}\\[1ex] 
\end{tabular}\label{table 1}}
\end{table}
\end{center}

\section{Important Experimental Gains From Signature Space Analysis}

This approach suggests \cite{E} some experimental issues that may mean
it is important to study these kinds of questions before analysis and
triggers are final, before there is data. By comparing inclusive
signatures some issues that affect separate measurements become less
important. Absolute measurements and beam luminosity questions
approximately drop out if one is only comparing collider $\sigma
\times BR.$ Even jet energy corrections that affect missing transverse
energy are of less importance when comparing channels than for
absolute measurements. It may be better to add all the ways to get a
given signature, with few cuts, to get large statistics. To fully take
advantage of these opportunities observables should be expressed as
ratios of event rates.

Even for different questions, inclusive signatures can be a good
approach.  For example, to search for the decay $t\rightarrow H^{+}b$ a
good method is to compare $\sigma \times BR$ for different apparent top
channels, since the effects of $H^{+}b$ occur only in certain channels
and not others. By avoiding cuts one can do even better.

Theorists need to work out the inclusive signature patterns of a
variety of approaches to supersymmetry breaking and for a variety of
underlying theories, filling in and extending the table (a larger
filled in table is shown in \cite{B}) . The approach can be extended
and sharpened a number of ways. \cite{B} suggests some approaches that
seem promising, and perhaps better options exist. One promising
extension \cite{E} is to work out where different theoretical
approaches lie on two (or three) dimensional plots, with numbers of
events of various inclusive signatures as the axes.  A given approach
with one set of parameters will be a point. When the parameters are
varied one will get a region around that point, a footprint.
Different approaches will lie in different regions of the plot. Even
with a single plot some approaches will be favored and some excluded
when there is data. When the analysis is done for a number of plots it
will be much more powerful in favoring some and excluding
others. Although we cannot uniquely determine predictions of an
underlying theory or a given way of breaking supersymmetry,
nevertheless the predictions of a given approach are typically
characteristic ones. Once there is data and theoretical analysis it
will be possible to zoom in on favored regions. In \cite{E} the full
region that a model such as mSUGRA occupies in signature space is
studied, allowing very sensitive tests of whether such an approach can
describe data (once there is data).

\section{CP Violation at Hadron Colliders --- Jet Charges}

The question of the origin(s) of CP violation is of unusual importance
in the effort to probe fundamental physics issues. We know today that
for confirmed laboratory experimental results a single phase from the
CKM matrix provides a satisfactory description. If the recent effects in
several hadronic channels arising from the $b\rightarrow s\bar{s}s$
penguin and studied by the Belle and Babar groups, who reported a
combined deviation from the Standard Model CP time dependent asymmetries
of 3.5$\sigma ,$ persist as data improves then another phase in addition
to the CKM phase must be present in the effective SSM
Lagrangian. Further, the size of the result implies it cannot be
described by a high scale effective theory that does not have
significant flavor dependence (in particular, the so-called mSUGRA
approach would be excluded, as would split supersymmetry). Further, the
baryon asymmetry of the universe requires CP violation and cannot be
described by the CKM phase alone.

If we knew the CP and phase structure of the effective 4D high scale
theory it could be extremely powerful in providing information about
how supersymmetry is broken, and about the compactified string
theory. Suppose for example we knew that all the laboratory CP
violation could arise from the CKM phase. Then all the F-term vevs
that parameterized supersymmetry breaking would be relatively real,
which would point to certain kinds of supersymmetry
breaking. Similarly it would point to certain kinds of
compactifications that did not allow relative non-zero phases.
Alternatively, if we knew from rare decay or collider data of other
phases that would exclude all approaches that did not allow them. A
very interesting question for theorists is to examine whether a single
phase in the underlying theory can emerge in the superpotential as the
CKM phase, and also emerge in the soft-breaking Lagrangian as a (say)
trilinear coupling phase, thus giving an apparently different phase to
the $b\rightarrow s\bar{s}s$ penguin.

Phases can be studied at b-factories, and in EDM experiments, which
makes the EDM experiments among the most important and fundamental
experiments possible for particle physics. Phases can also be studied
at colliders, which probe more and different phases as well as some of
the same ones as the EDM experiments. There has been considerable
analysis of possible CP studies at electron linear colliders. Since we
will have only hadron colliders for at least a decade after LHC begins
to take data, we need to learn how to study the phases at hadron
colliders. Some work has been done in that direction
\cite{F}. Magnitudes of phases can be studied in CP conserving
processes since they affect superpartner and higgs production rates
and decays, and signs of phases can be studied from non-zero triple
scalar products (one example is examined in \cite{F}).

Such studies will be much more powerful statistically if jet charges
can be measured, at least well enough to determine their signs. At LEP
jet charges were successfully measured, and presumably that will be
possible as well at hadron colliders for a large class of energetic
jets, but this has not yet been studied by experimenters. Note that at
LHC the initial state is in a sense CP even since most events
originate from gluons. Even if there are backgrounds because the
initial state of two protons, and the detector, are not CP-even, the
backgrounds should be fully calcuable. Whether one is searching for a
deviation from zero or a deviation from a known non-zero number is not
in principle different. SM processes will provide examples to test any
method. It will probably be sensible to use inclusive signatures here
too in order to increase statistics.

\section{Concluding Remarks}

When I give talks on the topics covered here I often get certain general
questions (in addition to technical ones), so let me address some of
them.  Are string theory and supersymmetry breaking too poorly
understood to carry out the sorts of studies I am advocating? I don't
think so --- a great deal has been learned over the past decade. While
no particular approach has emerged as favored, one can now go from a 10D
theory by a series of well defined steps, involving some assumptions
that can be realistic, to a construction that can make concrete
predictions for data. While doing that one learns a great deal about the
properties of the theory. The resulting phenomenology does depend on the
theory and the assumptions, which is good because once there is data
some approaches will be favored. I would go so far as to say that doing
such studies could greatly improve our understanding of string theory
and speed progress toward a Standard Supersymmetric String Model (SSSM)
--- and conversely.

From another point of view people say too little is known to make
progress. If so, that will not change when we learn nature is indeed
supersymmetric, since most workers already accept that. I think the
response here is that guessing the SM required very little data and
knowledge, and the same will likely be true for the SSM and SSSM. For
the SM it was known that there were V-A currents (rather than S,T,P),
chiral fermions (in modern language, parity violation in older
language), that weak interactions were weak, two neutrinos, the hadron
spectrum was known, and early deep inelastic scaling. On the
theoretical side gauge theories and the higgs mechanism and the
renormalizability of the electroweak theory and asymptotic freedom
were known. A coherent and correct picture quickly emerged. We may be
closer to such a ``phase transition'' in our understanding than most
people think.

Is there too much theory in the approach advocated here? Again,
examining recent history is suggestive. LEP provided a number of
precise measurements, and most importantly, three qualitative results:
that gauge couplings unified at a high scale in a supersymmetric world
with light superpartners, that there was an upper limit on the mass of
a fundamental higgs-like scalar (fundamental in that its couplings
were point-like on the LEP scale), and that whatever theory gave the
SM its foundations and extended it was weakly coupled because no LEP
measurement deviated from SM predictions at the level of a fraction of
a per cent. All these pointed to the supersymmetric extension. None of
them are apparent in the data alone.  Without a huge input from theory
LEP would have made no discoveries, but with it there are three major
discoveries. At LHC there will be major discoveries (such as the
existence of supersymmetry and higgs) that (only) require understanding
the SM ``backgrounds'', but going beyond them and interpreting the
data to learn what it implies will again require major theoretical and
experimental collaboration. In addition, for at least a decade theory
has to compensate for the lack of a linear collider.

One can ask why not just wait until there is data and then do the
sorts of analysis advocated here. In the text several technical
reasons were mentioned. Planning should guarantee that triggering and
cuts and selections are chosen in a coordinated way for all the
processes that might be compared. Arranging this, and trying to ensure
that jet charges can be measured so CP violation can be studied,
should be done before data is taken or opportunities may be partially
or fully lost. Thinking ahead of time about what may be found may lead
to analysis that will find non-obvious effects such as CP violation
and correlations between observables. And of course luck favors the
prepared mind. A crucial point is that \textit{patterns} of inclusive
signatures can replace the inability to measure the parameters of the
supersymmetry-breaking Lagrangian and the lack of a linear collider
during the LHC era, if we do the relevant experimental and theoretical
studies.

\section*{ACKNOWLEDGMENTS}

I appreciate very much discussions with Nima Arkami-Hamed, Frank Paige,
Pierre Binetruy, Joe Lykken, Tao Han, Steve Mrenna, Fabiola Gianotti,
Jake Bourjaily, Piyush Kumar, Maria Spiropulu, Peter Zerwas, and
particularly Brent Nelson, Liantao Wang, and Ting Wang.

\end{document}